\documentclass[twocolumn,showpacs,amsmath,amssymb,aps,prl,floatfix,nofootinbib,superscriptaddress]{revtex4}
\usepackage{graphicx}
\usepackage{dcolumn}
\usepackage{bm}

\begin{document}

\title{Analytic height correlation function of rough surfaces derived from light scattering}
\author{M. Zamani,$^1$ F. Shafiei,$^2$ S.~M. Fazeli,$^3$ M.C. Downer,$^2$  G.~R. Jafari,$^1$
\thanks{Email: g\_jafari@sbu.ac.ir} $^\dag$ \\
{\small $^1$Department of Physics, Shahid Beheshti University,
G.C., Evin, Tehran 19839, Iran} \\
{\small $^2$Department of Physics, The University of Texas at Austin, Austin, Texas 78712, USA}\\
{\small $^3$Department of Physics, University of Qom, Qom, Iran}
 }
\date{\today}

\begin{abstract}

We derive an analytic expression for the height correlation function of a rough surface based on the inverse wave scattering method of Kirchhoff theory. The expression directly relates the height correlation function to diffuse scattered intensity along a linear path at fixed polar angle. We test the solution by measuring the angular distribution of light scattered from rough silicon surfaces, and comparing extracted height correlation functions to those derived from atomic force microscopy (AFM).  The results agree closely with AFM over a wider range of roughness parameters than previous formulations of the inverse scattering problem, while relying less on large-angle scatter data.  Our expression thus provides an accurate analytical equation for the height correlation function of a wide range of surfaces based on measurements using a simple, fast experimental procedure.

\end{abstract}
\maketitle

%%%%%%%%%%%%%%%%%%%%%%%%%%%%%%%%%%%%%%%%%%%%%%%%%%%%%555
\section{I. Introduction}
Rough surfaces play an essential role in many physical phenomena including wave scattering \cite{Beckmann,Ogilvy,Voronovich},
friction \cite{Palasantzas}, adhesion \cite{Peressadko}, electrical conductivity \cite{Palasantzascond,ThinF}, capacitance \cite{Capac} and heat transport
\cite{Palasantzastermal,Volokitin}, and in applications ranging from
thin films to sensors \cite{Sangprb,Pradeep}.
The most direct way to measure morphology of rough surfaces is by scanning probe microscopy (SPM) \cite{Binnig,Solar}. However, probe size sometimes limits scanning resolution, in which case SPM images represent a
convolution of probe tip and intrinsic sample geometry \cite{Bogana,Buzio,Fazeli,Aue}. Light scatter, which is faster and less invasive than SPM, can also measure surface roughness via the inverse wave scattering method \cite{Jin,SIA}.
The spatial resolution of the scattering technique can easily be adjusted to a desired level by tuning the incident wavelength $\lambda$.

One of the most important parameters for describing rough surfaces is the height correlation function $Cor(R)=\langle h(x+R) h(x)\rangle / \sigma^2$, where $h(x)$ and $h(x+R)$ denote height above a mean surface height $\langle h \rangle = 0$ at horizontal positions $x$ and $x+R$, respectively, and $\sigma$ is the variance of $h$. The average is performed over $x$ for fixed $R$.
Other roughness parameters such as correlation length and roughness exponent can be derived from $Cor(R)$ \cite{Zhao2}.

Chandley \cite{Chandley} first showed that $Cor(R)$ could be obtained from a 2D Fourier transform of far-field scattered light intensity.
However, a time-consuming series of measurements of scattered intensity profiles for a wide range of incident angles was required. In Chandley's method, the height autocovariance function of the scattered wavefront is considered to be the same as the height autocovariance function of the rough surface.  This assumption, however, is only valid when wave vectors of incident and scattered light are close to the surface normal.  Moreover, Chandley's method was only practical for $\lambda\gg\sigma$.  Zhao \textit{et al.} \cite{Zhao} modified Chandley's method based on Kirchhoff's approximation and showed that $Cor(R)$ could be estimated rapidly from a \emph{single} measurement of scattered intensity along a linear detector array, for a single, arbitrary, fixed incident angle $\theta_1$ (see Fig.~1) and an arbitrary wavelength $\lambda$ without the restriction of $\lambda\gg\sigma$. Zamani \textit{et al.} \cite{Zamani,JMO} developed a rigorous mathematical foundation within Kirchhoff theory, using a saddle point approximation for calculating $Cor(R)$ of very rough surfaces from scattered light intensity measured along a special (in general curved) path along which both polar ($\theta_2$) and azimuthal ($\phi_2$) scattering angles (defined in Fig.~1) varied. However, Zamani's expression for $Cor(R)$ was not compared directly with that derived in Ref.~\cite{Zhao}, nor with experimental results. Chakrabarti \textit{et al.} \cite{Chakrabarti} developed an approach for obtaining $Cor(R)$ from the angular dependence of the mean differential reflection coefficient of a one-dimensional randomly rough dielectric surface, also in the Kirchhoff approximation.
\vspace{0.2cm}

In this paper, we derive a simple analytic relation between $Cor(R)$ and diffuse scattered intensity $\langle I_{diff} \rangle$ using a different mathematical approach than Ref.~\cite{Zhao}, although still in the framework of inverse scattering theory in the Kirchhoff approximation. Our expression relates $Cor(R)$ straightforwardly to $\langle I_{diff} \rangle$ in a simple experimental geometry:  fixed polar scattering angle $\theta_{2}$, measured as a function of azimuthal scattering angle $\phi_2$ between specular ($\phi_2 = 0$) and an arbitrary angle far from specular direction.
However, the derived relation between $Cor(R)$ and $\langle I_{diff} \rangle$ differs from that of Ref. \cite{Zhao}.
We demonstrate the accuracy of our expression by obtaining $Cor(R)$ from measured scattered intensity for surfaces whose correlation functions were obtained independently by atomic force microscopy (AFM),
and comparing the results with $Cor(R)$ obtained from the expression in Ref.~\cite{Zhao}.   The comparison shows that the present expression is less reliant on accurate, large-$\phi_2$, low-level scatter data, and thus avoids unphysical fluctuations at $R > 10\mu$m that can arise with the $Cor(R)$ expression of Ref.~\cite{Zhao} when the $\phi_2$ measurement range is restricted. Moreover, our optically extracted $Cor(R)$ agrees significantly better with AFM measurements than the Ref.~\cite{Zhao} expression for surfaces with high roughness ($\sigma \agt \lambda$).

\section{II. Height Correlation Function}
\subsection{A. Derivation}
In the following derivation, for simplicity, we restrict our attention to self-affine fractal rough surfaces --- \textit{i.e.} surfaces for which the height difference $\triangle=\langle|h(x_{1})-h(x_{2})|\rangle$ between two points $x_1$ and $x_2$ is related to their separation $R=|x_{1}-x_{2}|$ by a power-law $\Delta \sim R^{\alpha}$, where $\alpha$ is the roughness exponent \cite{barabasi}.   However, the general method is applicable to any homogeneous rough surface. The correlation function for self-affine surfaces can be written in the form $Cor(R)=\exp[-(R/\xi)^{2\alpha}]$, where $\xi$ is the correlation length $(Cor(R=\xi)=e^{-1})$ \cite{Zhao2} --- \textit{i.e.} the lateral length ($R$) at which the correlation function drops to $e^{-1}$ of its maximum at $R=0$. The value of $\alpha$ can be extracted from the structure function, which for a homogeneous isotropic rough surface is $H(R)=\left<[h(r)-h(r+R)]^{2}\right>$, and is related to the correlation function by $H(R)=2\sigma^{2}[1-Cor(R)]$ \cite{Zhao2}. For self-affine fractal rough surfaces the slope of the structure function on a log-log scale is equal to $2\alpha$ for $r<\xi$.
\vspace{0.2cm}

For a monochromatic incident wave
$\psi^{inc}(r)=e^{-i \mathbf{k}_{inc}\cdot \mathbf{r}}$ of wave vector $\mathbf{k}_{inc}$, where $\mathbf{r}$ represents position, the scattered wave in the Kirchhoff approximation is \cite{Ogilvy}
\begin{eqnarray}
 &\psi^{sc}(r) &= \frac{ike^{ikr}}{4\pi r}\int_{A_{M}}
(a\frac{\partial h}{\partial x_{0}}+b\frac{\partial h}{\partial
y_{0}}-c) \nonumber\\
& & \times ~e^{ik[Ax_{0}+By_{0}+ Ch(x_{0},y_{0})]}dx_{0} dy_{0},
\end{eqnarray}
where
\begin{eqnarray}
A&=&\sin \theta_{1}-\sin \theta_{2} \cos \phi_{2}, \nonumber \\
B&=&-\sin \theta_{2} \sin \phi_{2}, \nonumber \\
C&=&-(\cos \theta_{1} +\cos \theta_{2}), \nonumber \\
a&=&\sin \theta_{1}(1-R_{0})+\sin \theta_{2}\cos\phi_{2}(1+R_{0}), \nonumber \\
b&=&\sin \theta_{2} \sin \phi_{2}(1+R_{0}), \nonumber \\
c&=&\cos \theta_{2}(1+R_{0})-\cos \theta_{1}(1-R_{0}).
\end{eqnarray}
In Eq.~(1), $R_0$ is the reflection coefficient and the integral is over the mean reference plane $A_{M}$ of the rough surface.

\begin{figure}[t]
\includegraphics[width=9cm,height=6cm,angle=0]{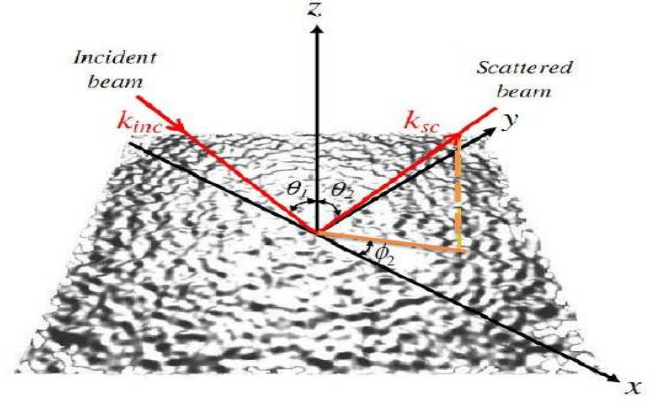}
\caption{Geometry for wave scattering from a rough surface.
} \label{fig1}
\end{figure}

The total scattered intensity includes coherent and diffuse parts. If
the spot size is much larger than the incident wavelength, the
coherent intensity $I_{coh}$ appears in the specular direction. The root mean square of surface heights can be calculated from \cite{Ogilvy}
\begin{eqnarray}
\label{sigma} \sigma=\frac{1}{kC}[\ln\frac{I_{0}}{I_{coh}}]^{\frac{1}{2}},
\end{eqnarray}
where $I_{coh}$ and $I_{0}$ are the scattered intensity from a rough and smooth surface, respectively, in the specular direction.
%The diffuse (non-specular) intensity $I_{diff}$ depends on the height correlation function.
For a surface with a Gaussian height probability distribution function of width $\sigma$, the mean diffuse intensity $\langle I_{diff}\rangle$ is related to $Cor(R)$ by \cite{Ogilvy}
\begin{eqnarray}
\label{Id} & \langle I_{diff}\rangle &=\langle\psi^{sc}\bar{\psi}^{sc}\rangle-\langle\psi^{sc}\rangle\langle\bar{\psi}^{sc}\rangle \nonumber\\
&&=\frac{k^{2}F^{2}}{2\pi r^{2}}A_{M}\exp(-g)\int_{0}^{\infty}[\exp(gCor(R))-1]\times \nonumber\\
&& J_{0}(kR\sqrt{A^{2}+B^{2}}) R dR.
\end{eqnarray}
Here the overbar denotes the complex conjugate of the scattered field $\psi^{sc}$, $F=\frac{1}{2}(\frac{Aa}{C}+\frac{Bb}{C}+c)$ depends on incident and scattered angles,
\begin{equation}
\label{g}
g=k^2\sigma^2(\mathrm{cos}\theta_1+\mathrm{cos}\theta_2)^2 =k^2\sigma^2C^2
\end{equation}
is the roughness criterion,
%$Cor(R)$ is height correlation function of rough surface.
and $J_{0}(kR\sqrt{A^{2}+B^{2}})$ is a zero-order Bessel function of the
first kind, which obeys the orthogonality relation \cite{Arfk}
\begin{eqnarray}
\label{arfken} \int_{0}^{\infty}J_{0}(UR)J_{0}(UR^\prime)U
dU=\frac{1}{R}\delta(R-R^\prime)
\end{eqnarray}
for an arbitrary function $U$, where
\begin{equation}
\label{main1}
U=k\sqrt{A^{2}+B^{2}}.
\end{equation}
for the scattering problem.
To eliminate the prefactor in Eq.~(\ref{Id}), we define normalized diffuse intensity
 $I_{d} \equiv \langle I_{diff}\rangle/\left(\frac{k^{2}F^{2}}{2\pi r^{2}}\right) e^{-g}A_{M}$.
Multiplying $I_d$ by $UJ_{0}(UR)$ and integrating yields
%\begin{eqnarray}
\begin{align}
\label{main}
\int^{\infty}_{0}& I_{d}  (U,g) J_{0}(U R)U dU = \int^{\infty}_{0}\int_{0}^{\infty}J_{0}(UR^{\prime}) \hspace{1.6cm} \nonumber\\
& ~~~~\times \left\{\exp[gCor(R^{\prime})]-1\right\}~J_{0}(U R) R^{\prime}dR^{\prime} U dU \nonumber\\
&= \exp[gCor(R)]-1,
\end{align}
%\end{eqnarray}
where we assumed \emph{constant} $g$ and used orthogonality relation (\ref{arfken}) to obtain the last expression.
Eq.~(\ref{main}) yields the desired analytic relation
\begin{eqnarray}
\label{cor1}
Cor(R)=\frac{1}{g}\ln{[\int^{\infty}_{0} I_{d}(U)J_{0}(U R)U dU+1]},
\end{eqnarray}
between $Cor(R)$ and scattered intensity.\\

\subsection{B. Discussion}
In (\ref{cor1}), $Cor(R)$ is most accurately determined when $I_d(U)$ is known over the entire range of $U$ values from zero to infinity.
However, in order to derive the analytic relation (\ref{cor1}), it was necessary to restrict the parameter $g$ to a constant value.  This in turn implies, for fixed $\theta_1$, a simple experimental set up with constant $\theta_2$.
Since $U$ and $g$ are functions of three common parameters ($k$, $\theta_{1}, \theta_{2}$), and $U$ also a function of $\phi_{2}$, fixing $g$ imposes some limitations on $U$.
Fixing $k$ (or equivalently $\lambda$) additionally limits the range of $U$.
$\sqrt{A^{2}+B^{2}}$ varies in magnitude from a minimum of zero to a maximum of 2.  Thus $U$ is restricted to the range $0 \rightarrow 2k$.
The scattering material, and available light sources and detectors, can further restrict the range of $k$, and thus $U$.
For example, metallic scatterers are highly reflective only for $k < \omega_p/c$, where $\omega_p$ is the plasma frequency.
Given such limitations, the question arises how accurately $Cor(R)$ can be determined in a given experimental scattering configuration.

Here we answer this question for an experimentally convenient scattering geometry with fixed $k$, $\sigma$, $\theta_1$ and $\theta_2$.  Then $g$ is fixed as required, and $U$ varies solely as a function of $\phi_2$. $I_d(U)$ can then be measured for selected fixed $\theta_2$ using a conventional rectangular charge-coupled device (CCD) array with its axes oriented along the $\theta_2$ and $\phi_2$ directions. Along the $\phi_2$ direction, $U$ then varies from $k\mid \sin\theta_{1}-\sin\theta_{2}\mid$ to $k[ \sin^2\theta_1 - 2\sin\theta_1\theta_2\cos\phi_2^{(max)} + \sin^2\theta_2]^{1/2}$ as $\phi_2$ varies from $0$ to $\phi_2^{(max)}$.  Choosing $\theta_1 = \theta_2$ ensures that values of $U$ down to zero are included.  Varying $\phi_2$ up to $\phi_2^{(max)} = \pi$ includes values of $U$ up to $k\mid\sin\theta_{1}+\sin\theta_{2}\mid$.  From Eq. (\ref{cor1}), measuring the scattered light intensity vs. $\phi_2$ then yields $Cor(R)$.

The correlation function formula in Ref.\cite{Chandley} and its modification in Ref.~\cite{Zhao} is derived from inverse Fourier transform of scattered intensity and given by,
\begin{eqnarray}
\label{zhao}
Cor(R)=1+\frac{1}{k_{\perp}\sigma}\ln[A\int I_{d}(k_{\perp},k_{||})e^{ik_{||}R}dk_{||}],
\end{eqnarray}
where, $k_{||}=k\sin(\phi_{2})$ and $k_{\perp}=k\cos(\theta_{2})$ and $A =1/\int I_{d}(k_{\perp},k_{||})dk_{||}$. Here the wave vector changed to $k_{||}$ parallel to the surface plays the role of U in our equation, and $\exp(ik_{||}r)$ is substituted by $J_{0}(UR)$ in our equation.
The main difference between these two equations arises at large $R$, where Eq.~(10) often generates unphysical oscillations in $Cor(R)$, as discussed below.
\begin{figure}[t]
\includegraphics[width=9cm,height=8cm]{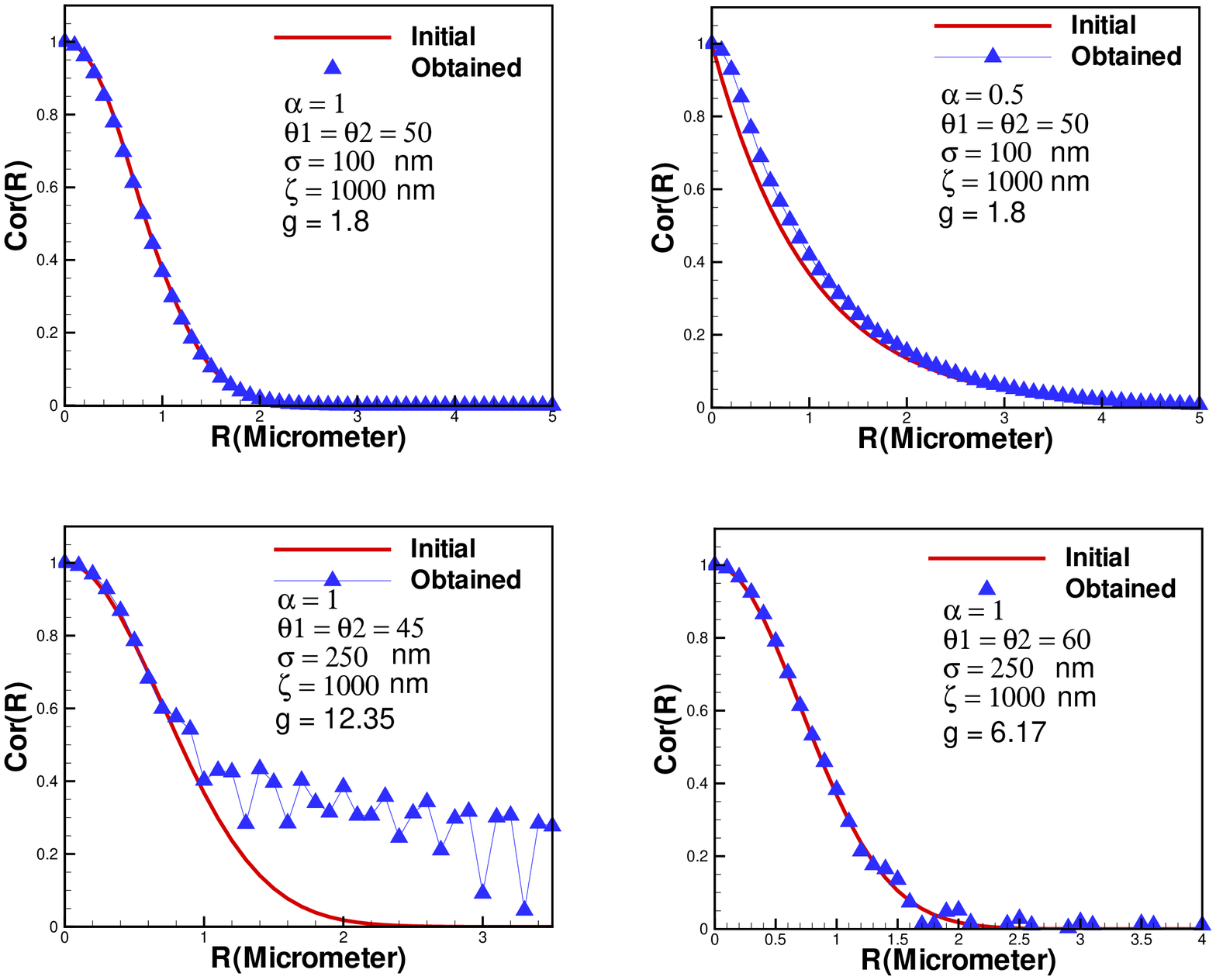}
\caption{Gaussian (Top-left) and exponential (Top-right) height correlation functions of surfaces with correlation length $\xi=1000$ nm, rms roughness $\sigma=100$ nm, $g=1.8$ calculated numerically from scattering of light of wavelength $\lambda=600$ nm, incident at $\theta_{1}=50^{\circ}$ (triangules) using the model described in the text, compared to the original height correlation functions (solid red curves). (Bottom-left) Deviation of correlation function from original one with increasing $\sigma$ and its compensation by increasing $\theta_{1}$,$\theta_{2} (Bottom-right).$ } \label{fig2}
\end{figure}

\subsection{C. Numerical Test}
To test our model numerically, we examine surfaces with two types of correlation functions (Gaussian and exponential) and various roughness ($\xi$, $\sigma$) and scattering ($\theta_1$, $\theta_2$, $\lambda$, $g$) parameters. Substituting these parameters in Eq.~\ref{Id} and inserting Gaussian or exponential correlation functions for $Cor(R)$, we calculate $I_{d}(U)$ in the Kirchhoff approximation. We then substitute this $I_{d}(U)$ in Eq.~\ref{cor1}. Fig.~\ref{fig2}(top-left and right) shows two calculated $Cor(R)$ curves, compared with the initially assumed one (red curves). The blue triangles show $Cor(R)$ calculated for a mildly rough surface $\xi=1000$ nm, $\sigma=100$ nm probed with scatter parameters $g=1.8$, $\theta_1=\theta_2=50^{\circ}$, $\lambda=600 nm$.  In this case the calculated $Cor(R)$ agrees very closely with the initially assumed one:  coefficient of determination $r^2$ is 0.99 for both correlation functions, and the fit standard deviation error std $=0.004$ and 0.02, respectively, for Gaussian and exponential functions. As $\sigma$ increases relative to $\lambda$, the calculated $Cor(R)$ increasingly deviates from the initial assumption (Fig.~\ref{fig2}, bottom left).  Reducing $g$ by increasing $\theta_1$, $\theta_2$ compensates for large $k\sigma$, and recovers good agreement (Fig.~\ref{fig2}, bottom right). Generally, keeping $g < 8$ yields good agreement, as expected for the Kirchhoff approximation, which is not accurate for $g\gg1$.
\vspace{0.2cm}

\section{III. Experiments}
\subsection{A. Procedure}
\begin{figure}[t]
\includegraphics[width=8cm,
height=4.5cm]{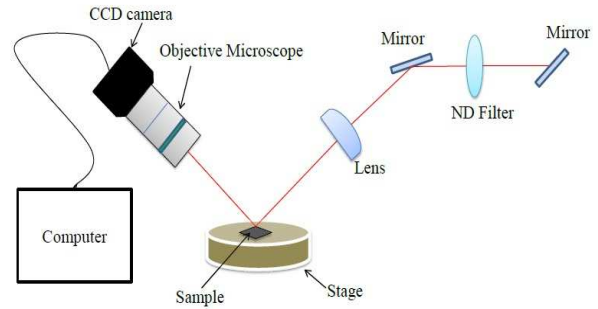}
\caption{Experimental setup for measuring intensity profile of light scattered from a rough surface.  Light source is a HeNe laser; steering mirrors enable adjustment of incident angle $\theta_1$; ND filter ensures detector remains unsaturated; CCD camera is centered on the specular reflection angle, and located in the focal plane of a microscope objective collecting lens.} \label{fig3}
\end{figure}

To test the model experimentally, we used the light scattering set up shown in Fig.~\ref{fig3}.  A lens ($f = 5$ cm) focused a He-Ne laser beam ($\lambda=633nm$, power $\sim 1$ mW, p-polarization) at $45^\circ$ incidence angle to spot size $w_o \approx 20 \mu$m onto unpolished back sides of commercial silicon wafers that were mounted on a translation stage, to enable convenient probing of multiple spots on each surface.  A microscope objective of numerical aperture $0.42$ collected scattered light without polarization discrimination in a cone of $25^{\circ}$ half-angle around the specular direction.  A charge-coupled device (CCD) camera placed in its focal plane, centered at $\theta_2 = 45^\circ$, $\phi_2 = 0^\circ$, recorded the intensity profile of scattered light over angular ranges $20^\circ < \theta_2 < 70^\circ$, $-25^\circ < \phi_2 < 25^\circ$.  All light detectable above noise was scattered within this cone angle, so there was no need to change the position of the CCD to collect light over a wider range.

\subsection{B. Results and Analysis}
Fig.~4a shows a typical measured 2D scattered intensity profile obtained in this configuration with 0.3 ms exposure.  However, exposures as long as 30 ms were used to record low level scatter at large $\phi_2$.  Fig.~4b shows a line-out along the azimuthal direction passing through the profile at the specular polar angle $\theta_2 =  45^\circ$.  The coherent specular peak and diffuse scattered light are both evident in this line-out.  For data used in quantitative analysis of surface roughness, it was important to ensure that the detector remained unsaturated over the entire dynamic range of scattered light intensity.  To this end, we recorded each scattered intensity profile several times with different neutral density filters inserted in the path of the incident beam, so that the intense coherent peak and the weak tails of the diffuse profile were both recorded within the CCD's linear response range. We then assembled the composite profile from the separate recordings.  The shape of the scattered intensity profile, and extracted roughness parameters, did not depend significantly on polarization of the incident beam.  From the measured linearized total scattered intensity profile $I(\phi_2)$, we separately fitted the coherent central peak $I_{coh}(\phi_2)$ and diffuse profile $I_d(\phi_2)$, the latter from the edge of the central coherent peak out to the largest angle $\phi_2$ at which scattered light was detectable above noise.  Since samples were mounted on a translation stage, we measured scattered profiles at many different spots on the surface to determine the statistical variance of key roughness parameters.  We then calculated $\sigma$ from Eq.~(\ref{sigma}), and $Cor(R)$ from Eq.~(\ref{cor1}) or (\ref{zhao}). Finally we determined $\sigma$ and $Cor(R)$ independently from AFM images, such as the one shown in Fig.~4c, from the same regions of the surface from which light was scattered.

\vspace{0.2cm}

\begin{figure}[t]
\includegraphics[width=8.8cm,height=6.7cm,angle=0]{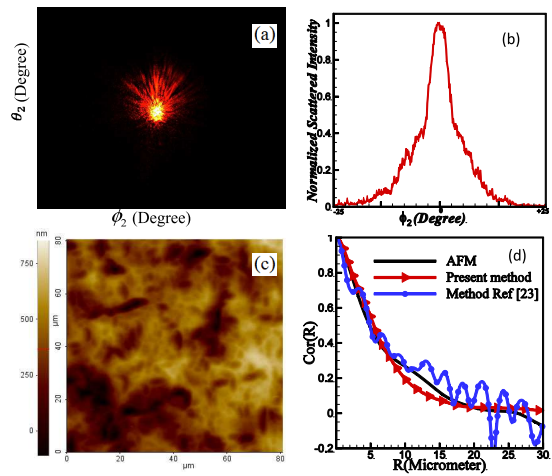}
\caption{Roughness data for Si surface with $\sigma =175$ nm $\sim \lambda/4$.  (a) 2D scattered intensity profile recorded with CCD camera centered in the specular direction $\theta_2 = \theta_1 = 45^\circ$ and $\phi_2 = 0$.   (b)Line-out of the profile in panel (a) along the azimuthal ($\phi_2$) direction, passing through the specular polar angle $\theta_2 = 45^\circ$, showing coherent specular peak centered at $\phi_2 = 0$, and diffuse scattered profile at larger $\phi_2$. (c) AFM image of the silicon surface. (d) correlation function for the silicon sample obtained from present method (red-triangle), AFM (black) and the method of Ref.~\cite{Zhao} (blue-circle).
 } \label{fig4}
\end{figure}

%%%%%%%%%%%%%%%%%%%%%%%%%%%%%%%%%%%%%%%%%%%%%%%55
%%%%%%%%%%%%%%%%%%%%%%%%%%%%%%%%%%%%%%%%%%%%%%%%%%%%%%5
%%%%%%%%%%%%%%%%%%%%%%%%%%%%%%%%%%%%%%%%%%%%%%%%%%%%%%%

\begin {table}[t]
\caption {Statistical surface roughness parameters obtained from AFM, and from light scattering using the present method and the method of Ref.~\cite{Zhao}.} \label{tab:Table Title}
\begin{center}
    \begin{tabular}{  | l | l | p{2.5cm} | p{2.5cm} |}
    \hline
        &   AFM & Scattering (present method)& Scattering (Ref.~\cite{Zhao} method)\\ \hline
    $\sigma(\mu m)$ & $0.175\pm0.002$ & $0.167\pm0.002$ & $0.167\pm 0.002$  \\ \hline
    $\xi(\mu m)$ & $6.9\pm0.2$ & $7.1\pm0.2$ & $6.8\pm0.5$ \\ \hline
    $\alpha$ & $0.73\pm0.02$ & $0.81\pm0.02$ & $0.83\pm0.03$\\ %\hline
%    c & $0.152\pm0.003$ & $0.177\pm0.001$ \\ \hline
%    d & $0.337\pm0.003$ & $0.671\pm0.002$ \\
     \hline
    \end{tabular}

\end{center}
\end {table}

\begin{figure}[t]
\includegraphics[%trim=2.5cm 9cm 0 3cm,
width=6cm,height=5cm]{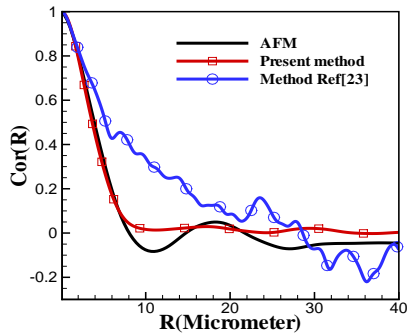}
\caption{$Cor(R)$ for Si surface with $\sigma=657 nm \approx \lambda$, derived with present method (red-triangle), AFM (black), and the method of Ref.~\cite{Zhao} (blue-circle).
 } \label{fig5}
\end{figure}

%[\textit{For calculating $\sigma$ with inverse scattering method the very smooth side of silicon surface is used}.]
Table $I$ shows extracted $\sigma$, $\xi$ and $\alpha$ values from AFM, and from light scattering using the present method and the method of Ref.~\cite{Zhao}, for Si surfaces with $\sigma \sim \lambda/4$.   The uncertainties represent standard deviations of measurements acquired from a large number of spots on different, but nominally identical, Si samples.  Roughness parameters obtained from the two light scattering methods agree within the stated uncertainties.   The light-scatter value of $\sigma$ is $\sim 5\%$ smaller, and of $\alpha$ $\sim 10\%$ larger, than the corresponding AFM values, while all $\xi$ values agree within measurement uncertainty.

\vspace{0.2cm}

Fig.~4d compares $Cor(R)$ functions obtained from AFM (black curve), and from light scatter analysis using Eq.~(\ref{cor1}) (red) and Eq.~(\ref{zhao}) (blue), for the same surfaces used for the data in Table I.  Three significant differences between the two light scatter analyses now emerge.  First, the red curve, extracted from scatter profiles with maximum exposure 0.3 ms, did not change perceptibly when high-$\phi_2$ data from profiles with 30 ms exposure were included.  In contrast, the blue curve, extracted from profiles with exposure up to 30 ms, changed substantially when limited to data acquired with only 0.3 ms exposure.  Thus, evidently, our Eq.~(\ref{cor1}) relies much less on accurate, large-angle, low-level scatter data than Eq.~(\ref{zhao}).  Second, the red curve agrees somewhat better with the AFM (black) curve ($r^{2}$=0.97, standard deviation std=0.03) than the blue curve ($r^{2}$ =0.89, std=0.11).  Third, large fluctuations appear in the blue curve at large $R$ that are not present in the other $Cor(R)$ curves.  We attribute these fluctuations to truncation of large $k_{||}$, and thus large $\phi_2$, values when evaluating the integral in Eq.~(\ref{zhao}).   In principle, these fluctuations could be reduced by using a wider CCD camera, or by translating the CCD in the azimuthal direction and using long exposures to acquire data over a wider $\phi_2$ range. Taken together, these differences demonstrate that Eq.~(\ref{cor1}) can yield a more accurate $Cor(R)$ than Eq.~(\ref{zhao}) when only low-exposure, low-azimuthal-angle data is available.

Further differences between Eqs.~(\ref{cor1}) and (\ref{zhao}) emerge when analyzing light scatter data from very rough ($\sigma \agt \lambda$) surfaces.
As an example, Fig.~5 shows $Cor(R)$ functions extracted from a Si surface with $\sigma=657 nm \approx \lambda$.  The two light scatter analyses now diverge significantly from each other, the curve based on Eq.~(\ref{cor1}) (red) agreeing much more closely ($r^{2}$ =0.98 and std=0.03) with $Cor(R)$ obtained from AFM (black curve) than the curve based on Eq.~(\ref{zhao}) (blue). Thus Eq.~(\ref{cor1}) appears to yield more accurate $Cor(R)$ than Eq.~(\ref{zhao}) for very rough surfaces.

\vspace{0.2cm}
\section{IV. Conclusion}
In summary, we introduce a method for calculating the height correlation function $Cor(R)$ of a homogeneous rough surface in the framework of Kirchhoff theory that yields better agreement with AFM measurements over a wider range of roughness parameters than previous formulations.  Test measurements use a simple experimental geometry with a single CCD camera at a fixed polar scattering angle that is amenable to time-resolved measurements.   The extracted $Cor(R)$ is free of fluctuations at large $R$ that arise with previous analysis methods of comparable simplicity.

\vspace{0.2cm}
\section{Acknowledgments}
 This work was supported by the Robert Welch Foundation (grant F-1038).  We thank Dr Xiaoqin Li for the use of an AFM system.
\\

%@@@@@@@@@@@@@@@@@@@@@@@@@@@@@@@@@@@@@@@@@@@@@@@@@@@@@@@@
%@@@@@@@@@@@@@@@@@@@@@@@@@@@@@@@@@@@@@@@@@@@@@@@@@@@@@@@@

\end{document}